\documentclass{article}

\usepackage[english]{babel}

\usepackage[letterpaper,top=2cm,bottom=2cm,left=3cm,right=3cm,marginparwidth=1.75cm]{geometry}

\usepackage{amsmath}
\usepackage{graphicx}
\usepackage[colorlinks=true, allcolors=blue]{hyperref}

\title{Resolved frustrated tunneling ionization in asymmetrical fast oscillation of above-threshold ionization spectrum}
\author{Lifeng Wang$^1$$^+$, Hao Teng$^1$$^,$ $^3$*, Fei Li$^1$, Bingbing Wang$^1$$^,$ $^2$$^,$ $^3$*, Xiaoxin Zhou$^4$,\\
Peng He$^3$,and Zhiyi Wei$^1$$^,$ $^2$$^,$ $^3$*\\
$^1$Beijing National Laboratory for Condensed Matter Physics, Institute of Physics, \\
Chinese Academy of Sciences, Beijing 100190, China.\\
$^2$University of Chinese Academy of Sciences, Beijing 100049, China.\\
$^3$Songshan Lake Materials Laboratory, Dongguan 523808, China.\\
$^4$College of Physics and Electronic Engineering, Northwest Normal University, \\
Lanzhou, 730070, China.\\
$^+$Currently with Institute for the Frontier of Attosecond Science and Technology (iFAST), \\CREOL and Department of Physics, University of Central Florida,\\ 4111 Libra Drive, Orlando, FL 32816, USA.\\
*Corresponding authors: hteng@iphy.ac.cn, wbb@aphy.iphy.ac.cn, zywei@iphy.ac.cn.}

\begin{document}
\maketitle

\begin{abstract}
Tunneling ionization is one of the fundamental electron dynamics, which has wide applications in ultrafast physics \cite{hutten2018ultrafast,yun2018coherent,eichmann2009acceleration}.  When frustrated tunneling ionization (FTI) is considered \cite{eilzer2014steering}, the tunneling rate is not equivalent to ionization rate. However, it is hard to resolve the effects of FTI and direct tunneling ionization (DTI) in ionization spectrum experimentally. Here we report the first observation of the asymmetrical fast oscillation in above-threshold ionization (ATI) spectrum of Argon as function of carrier-envelope phase (CEP), to the best of our knowledge. Simulation results identify that in the experimental ATI spectrum, the $\pi/5$ oscillation originates from the quantum interference of electrons in FTI, while DTI is responsible for the asymmetry. Our results provide clear evidence to resolve the effects of direct tunneling and FTI in a new physical regime.
\end{abstract}

\section{Introduction}

When atoms or molecules are exposed to intense ultrafast lasers, the tunneling ionization of electron is a fundamental process for several physical phenomenon, such as ATI, filamentation and high-order harmonic generation (HHG) \cite{ren2019line,wang2021two,boltaev2020application}. Commonly, the tunneling rates is equal to ionization rate. However, this is not true if the Coulomb potential is considered. Attracted by the Coulomb potential, the electron has possibility to have high negative energy at the tunnel exit and relaxes to Rydberg state (RS), this process is dubbed as frustrated tunneling ionization\cite{nubbemeyer2008strong,liu2021electron}. The RS atoms or molecules has been proved to have wide applications \cite{eichmann2013observing,dunning2009engineering,chini2014coherent}. There are two mechanisms of the formation of RS atoms or molecules, which are still under debate. One is that RS are populated by multiphoton resonance\cite{volkova2011ionization}. Another one considers that the electron is recaptured into RS after tunneling ionization\cite{li2014fine}. From the experimental point of view, the formation of RS is usually investigated in terms of laser intensity with multicycle femtosecond laser pulses\cite{lv2016comparative,xu2020observation}. However, FTI happens with direct tunneling ionization (DTI) at the same time. How to simultaneously demonstrate the contributions of both ionization processes to the final ionization probability is still an open question.

For ultrashort laser pulses, especially in few-cycle regime, there is an important parameter CEP, which has improved to be critical in many physical phenomena\cite{milovsevic2006above,wang2016carrier,ayuso2021ultrafast}. However, the CEP effect on the formation of RS atoms or molecules has not been extensively investigated so far. T. Nakajima \emph{et al.}\cite{nakajima2006phase} has theoretically investigate the CEP dependent RS atoms in the multiphoton ionization regime, but without experimental evidence. H. Yun \emph{et al.}\cite{yun2018coherent} has observed HHG from Hydrogen by 5 fs laser pulses. The authors contributed the coherent extreme-ultraviolet emission to the excited atoms through the frustrated tunneling ionization and the population of excited states is calculated in terms of CEP.J. Liu \emph{et al.}\cite{liu2020dynamics} theoretically optimize the CEP for the creation of RS.A. Gürtler \emph{et al.}\cite{gurtler2004asymmetry} experimentally investigate the direction of electron ejected from RS rubidium with few-cycle radio-frequency pulses in multiphoton ionization regime. The authors have found that the electron ejection direction strongly depends on CEP.

Recently, M Kübel \emph{et al.}\cite{kubel2021high} have experimentally found that the modulation periodicity of ATI spectrum from Cs in gas-phase driven by a 3.1 $\mu$m laser is $\pi/3$. The fast oscillation happens in the energy range of [5Up, 7Up] in the experiment, where Up is the ponderomotive energy. The authors explain that the unusual behavior originates from the interference of few recollision quantum orbits of the ionized electron, which is independent with RS excitation process.

Here we report the first experimental evidence to distinct the effects of FTI and DTI simultaneously in a new physical regime. The ATI spectrum for Argon is measured driven by a few-cycle laser pulse. It is observed a modulation of whole ATI photoelectron spectrum with a period of $\pi$/5 for relative CEP=[0, $\pi$], which is the fastest CEP-induced modulation period, to the best of our knowledge. Interestingly, this modulation decreases dramatically in the range of [$\pi$, 2$\pi$]. Our simulations results, which are well agreed with experimental data quantitatively, identify the importance of two kinds of ionization processes. The fast oscillation in ATI spectrum originated from the interferences of electrons in RS during FTI. While the asymmetry comes from the electrons DTI in the few-cycle laser fields. Our results are direct evidence to resolve the different effects of FTI and DTI in ATI spectrum in a new physical regime. Our findings help to understand the mechanism of formation of RS atoms in strong laser field and shed light on a new way to control the electron dynamics in atoms, which can be further developed in molecules and solids. 

\section{Experiment}
Our experiment is performed on a system of isolated attosecond pulse measurement in Institute of Physics, Chinese Academy of Sciences \cite{zhong2013effects}. A commercial Ti: sapphire chirped-pulse amplifier (Femtolaser Compact PRO) provides laser pulses with central wavelength of 760 nm, pulse duration of 25 fs, pulse energy of 0.8 mJ laser pulses at 1 kHz repetition rate. The laser spectrum is broadened in a differentially pumped neon-filled hollow-core fiber and then temporally compressed by chirped mirrors\cite{robinson2006generation}. The whole spectrum range covers from 550 nm to 950 nm. The CEP of the laser pulse is actively stabilized with a fluctuation of 80 mrad in more than 7 hours by locking both the oscillator and the amplifier loops. Finally, laser pulses with energy of 0.25 mJ and duration around 7 fs FWHM are obtained, which is around two optical cycles\cite{zhang2014long}. In our setup, the CEP of the 7 fs laser pulse is tuned by a pair of fused silica wedges in the optical beam. 

The linearly polarized laser pulses are focused into a vacuum chamber with backing vacuum pressure of $10^{-7}$ mbar. A motorized aperture is set to control the laser power, which is important in our experiment. The IR intensity at focus is controlled by the motorized aperture from $10^{11} W/cm^{2}$ to $10^{14} W/cm^{2}$. In the range of focus, Argon atoms are injected by a gas jet to interact with IR laser to trigger ATI process. The photoelectron spectrum for ATI is recorded with a time-of-flight (TOF) spectrometer in terms of the CEP of input few-cycle IR laser (see Supplemental material for details of experimental setup and procedures).

\begin{figure}[htbp]
\centering\includegraphics[width=12cm]{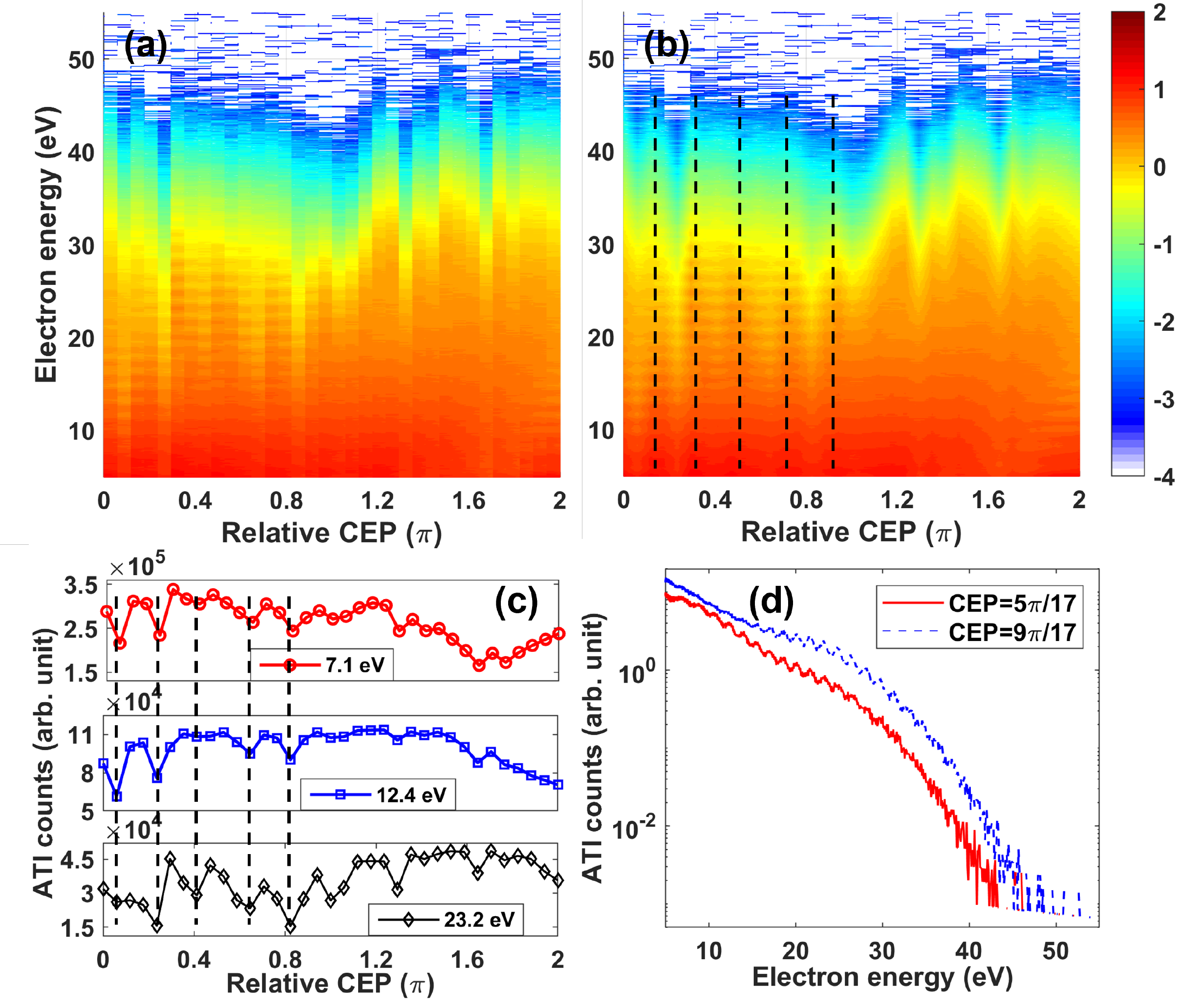}
\caption{The measured ATI spectrogram of Argon driven by few-cycle laser pulses for (a) the original data and (b) the fitting data, which shows a clear oscillation of $\pi/5$. (c) The ATI spectra for three different energies in terms of CEP. (d) The ATI spectra for two different CEPs.}
\label{fig:Figure1Exp}
\end{figure}

Measured photoelectron energy spectra for ATI in terms of CEP are shown in Fig. \ref{fig:Figure1Exp} (a), in which CEP is tuned with a step of $\pi/17$, which is calibrated by HHG spectrum\cite{ye2014full}. The measurement range of the TOF is 5 to 200 eV. There are several interesting observations in the experimental results. Firstly, clear oscillation in ATI spectrum is found with a period of $\pi/5$ in the range of [0, $\pi$] after data fitting, as shown in Fig. \ref{fig:Figure1Exp} (b). The Nyquist frequency of the CEP scan is $8.5/\pi$, which is higher than the frequency we measured. Secondly, another striking feature of our experimental results is that the oscillation decreases dramatically in the range of [$\pi$, $2\pi$]. Compared to experimental results in Ref.\cite{kubel2021high}, the CEP-dependent oscillation repeats in a range of [0, $4\pi$]. The differences between the observations from Ref.\cite{kubel2021high} and our experiment may indicate that there are different underlying mechanisms.

To see our experimental finding more clearly, the measured photon electron counts at three different energy ranges (±0.8 eV) is shown in Fig. \ref{fig:Figure1Exp} (c). The oscillation for 7.1 eV shows a larger modulation depth than the one for 12.4 eV. The modulation of ATI counts for 23.2 eV is most significant in range of [0, $\pi$]. Another interesting observation is that the suppression of ATI spectrum in full spectral range, as shown in Fig. \ref{fig:Figure1Exp} (d). For relative CEP=5$\pi$/17, the ATI spectral (solid red line) is suppressed in the whole energy range of [5, 40] eV, compared with the spectrum when CEP=9$\pi$/17. The origin for this observation will be discussed below.
\section{Simulation results and discussion}
In order to understand the experimental results, the one-dimensional time-dependent Schrödinger equation (TDSE) is used to describe the interaction between the atom and the linearly polarized laser pulses. The electric field of the laser pulses which is expressed as

\begin{equation}
E(t)=\sum_{i}E_if_i(t)sin(\omega_it+\phi_i)
\end{equation}
where $E_i$ is the electric field amplitude, $f_i(t)=sin^2(\pi t+\phi _i)$ is the electric field envelope, $\omega _i$ is the frequency of the laser pulse, and $\phi _i$ is the CEP of the laser pulse. The initial state is obtained by diagonalizing the field-free Hamiltonian $H_0$ of the atom and the TDSE is solved by the split-operator method. Hence, the population of the bound states after the end of laser pulse is obtained by projecting the final wave function to the eigenstate of $H_0$ (see Supplemental material for details of theory).

\begin{figure}[htbp]
\centering\includegraphics[width=12cm]{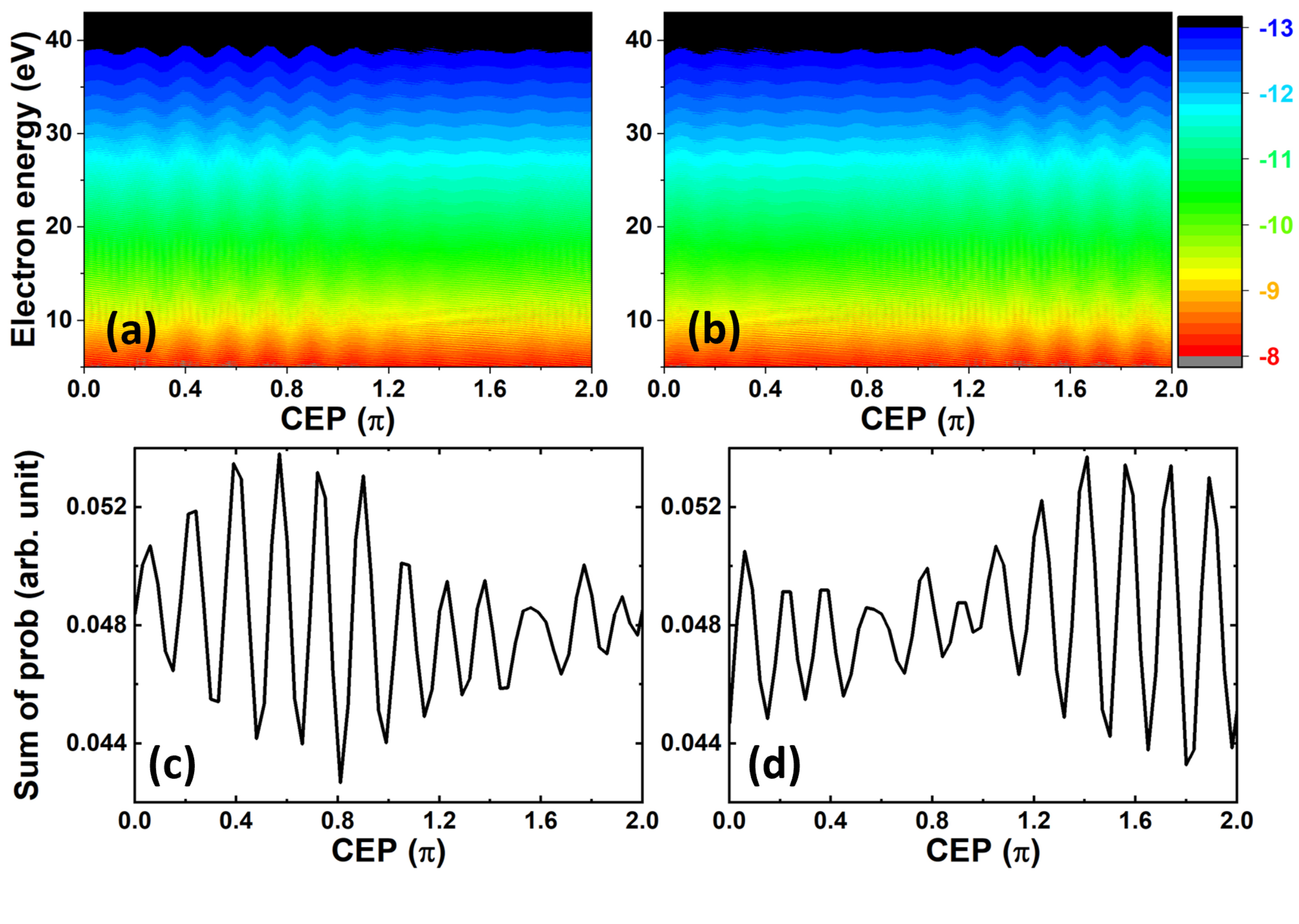}
\caption{Calculated ATI spectra vs CEP along the positive (a) and negative (b) directions of laser’s polarization. Summation of ionization probability vs CEP along the positive (c) and negative (d) directions of laser’s polarization.}
\label{fig:Figure2CaATI}
\end{figure}

Our simulations clearly reproduce the main characterizations in experimental results. The simulation results for Argon atom are shown in Fig. \ref{fig:Figure2CaATI} (a) with calculation parameters of 660 nm, 760 nm and 860 nm with same CEP (three-color laser field as represented in Eq. (1)), 7.8 fs and peak intensity $1.9 \times 10^{14} W/cm^{2}$. Because there is one TOF spectrometer in our experiment along laser’s polarization direction, only electrons along one direction are summed in the calculation. In the range of [0, $\pi$], there is oscillation of ATI spectrum with a period of $\pi/5$ and the oscillation decrease obviously in the range of [$\pi$, $2\pi$]. Furthermore, if the electrons along another direction are summed as shown in Fig. 2 (c), which is not measured in experiment, the oscillation appears in the range of [$\pi$, $2\pi$]. To see this oscillation more clearly, the total ionization probability along two directions in terms of CEP are shown in Fig. 2 (c) and (d) with the energy of the electron in region of [5, 40] eV. The remarkable agreement between the simulation results and experimental data give rise to a question: what is the underlying mechanism of these experimental observations?

\begin{figure}[htbp]
\centering\includegraphics[width=12cm]{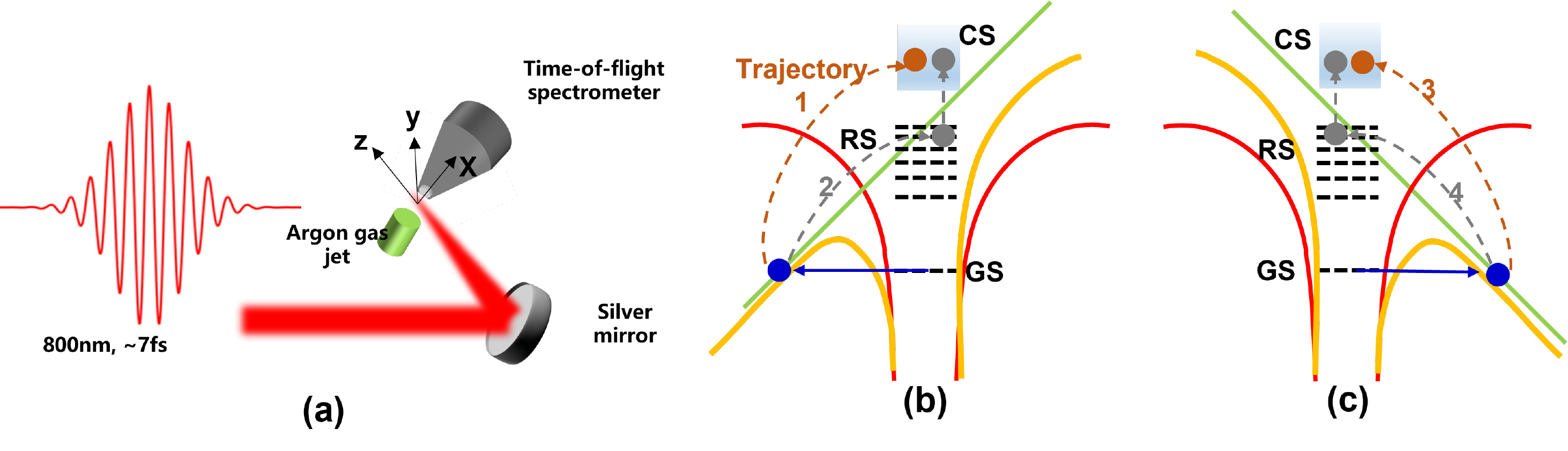}
\caption{(a) The schematic of simplified experimental setup. (b) Two trajectories of electrons ionization along +x direction. (c) Two trajectories of electrons ionization along -x direction. GS: ground state, CS: continuum state and RS: Rydberg state. The red, green and yellow lines are the Coulomb potential, the external laser field and the total electric field, respectively in (b) and (c).}
\label{fig:Figure3PhysPic}
\end{figure}

Here, we propose a semi-classical picture of electron ionization, as illustrated in Fig. \ref{fig:Figure3PhysPic}. We simplify the experimental setup, as shown in Fig. \ref{fig:Figure3PhysPic} (a). The ionization of electron is discussed under two conditions. When the electric field oscillates along +x direction, as shown in Fig. \ref{fig:Figure3PhysPic} (b). The electron is firstly tunneling ionized from ground state directly, which is labeled as trajectory 1. Then the electron is accelerated by the electric field and has possibility to be rescattered from ion. On the other hand, the electron has possibility to be captured by Coulomb potential into RS during tunneling. Later, the electron may be further ionized from RS to continue state by laser field, which is labeled as the trajectory 2. The ionized electrons from both trajectories 1 and 2 are measured in experiment. When the electric field oscillates along -x direction, as shown in Fig. \ref{fig:Figure3PhysPic} (c).  There are another two trajectories, similar to Fig. \ref{fig:Figure3PhysPic} (b). For trajectories 1 and 3, the electrons move along +x and -x directions, which depend on the electric field. However, since the Coulomb potential is spatially symmetry, both trajectories 2 and 4 may contribute to a certain RS, and the ionization from RS is independent with the direction of laser’s polarization. These differences have important role in our experimental results, as discussed in below.

\begin{figure}[htbp]
\centering\includegraphics[width=12cm]{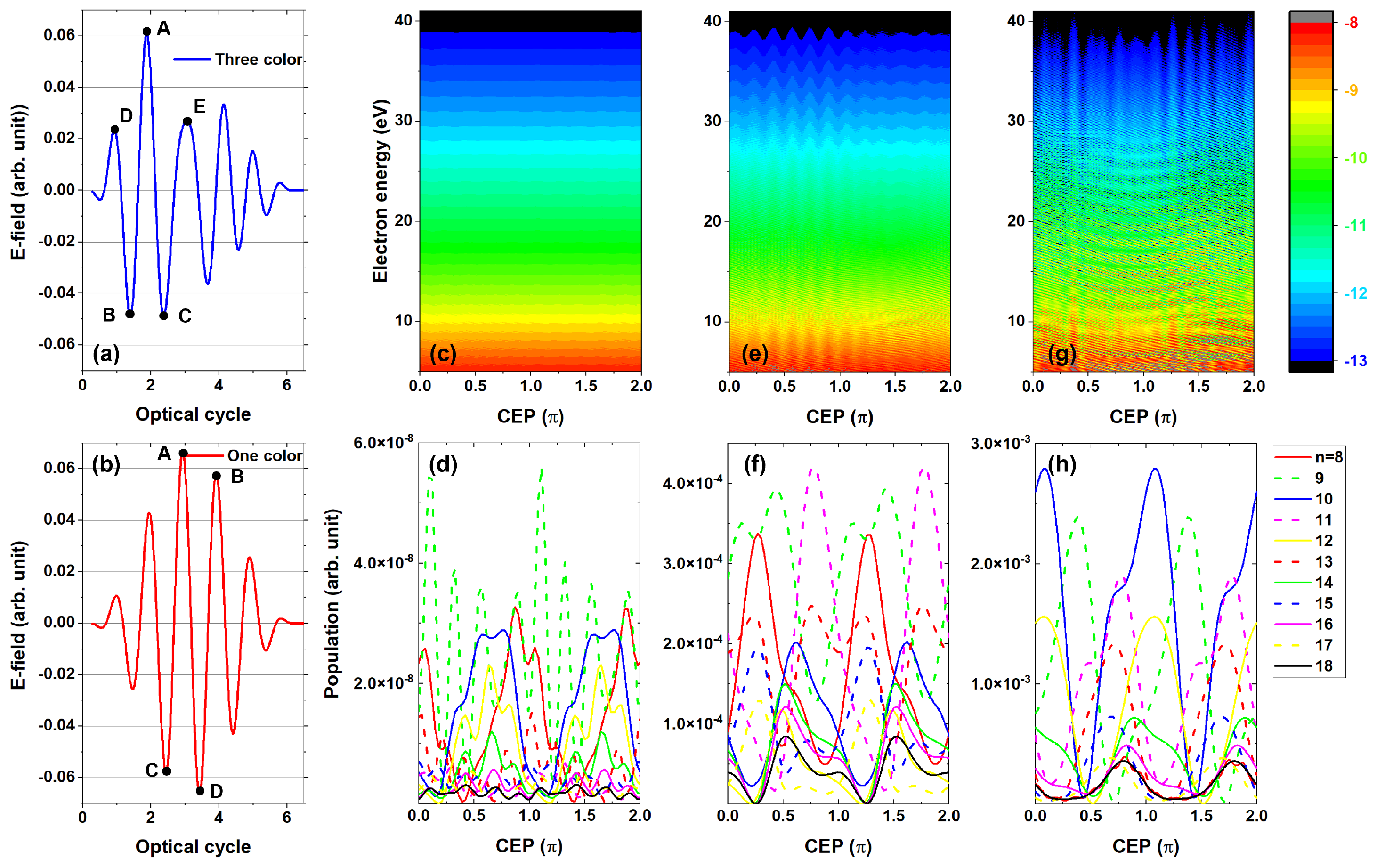}
\caption{Electric filed with one-color (a) and three-color (b) driving laser pulse. Calculated ATI spectrum and corresponding population of RS with three-color driving laser field at intensity of $4 \times 10^{13} W/cm^{2}$ (c and d), and intensity of $1.9 \times 10^{14} W/cm^{2}$ (e and f). Calculated ATI spectrum and corresponding population of RS with one-color driving laser field at intensity of $1.4 \times 10^{14} W/cm^{2}$ (g and h).}
\label{fig:Figure4ATIandRS}
\end{figure}

Comparing with experimental data, two questions need to be answered: 1. why the oscillation has a period of $\pi/5$ in range of relative CEP=[0, $\pi$]? 2. Why this oscillation dramatically decreases in the range of relative CEP=[$\pi$, $2\pi$]? Firstly, we investigate the laser intensity effect on the ATI spectrum driven by three-color (660 nm, 760 nm and 860 nm with relative amplitude of electric field 0.6:1:0.6) laser field, as shown in the Fig. \ref{fig:Figure4ATIandRS} (a). To understand the oscillation period, the population of atoms in RS are also calculated. When the laser intensity is $4 \times 10^{13} W/cm^{2}$, the electrons are mostly ionized by multiphoton ionization. In multiphoton ionization regime, the ionization probability is independent to the population of the RS. Therefore, although the population for the principle number of odd and even RS oscillate with CEP, as shown in Fig. \ref{fig:Figure4ATIandRS} (d), we cannot observe any oscillation in the ATI spectrum, as shown in Fig. \ref{fig:Figure4ATIandRS} (c).  However, when the laser intensity increases to $1.9 \times 10^{14} W/cm^{2}$, the DTI dominate the electron ionization process. Electrons from trajectories 1 and 2 start to play an important role. The populations of RS increase four orders of magnitude ($\sim10^{-4}$), as shown in Fig. \ref{fig:Figure4ATIandRS} (f). More importantly, the CEP dependent odd and neighboring even RS has a phase shift around $\pi/4$. By using different absorbing boundaries in calculation, one may find that the contribution mainly comes from RS with principle number larger than 7 (see Supplemental material). Consequently, the total RS has an oscillation of $\pi/5$ by the interference coming from the electrons of trajectory 2. The higher population of RS, the lower population of ionized electron. Therefore, the calculated ATI spectrum has an oscillation of $\pi/5$, as shown in Fig. \ref{fig:Figure4ATIandRS} (e) and the whole spectrum is suppressed for different CEPs, as shown in Fig \ref{fig:Figure1Exp} (d).

The population of RS is symmetry for CEP=[0, $\pi$] and [$\pi$, $2\pi$], as shown in Fig. \ref{fig:Figure4ATIandRS} (f). However, strong asymmetrical ATI spectrum appears in Fig. \ref{fig:Figure4ATIandRS} (e).  In order to explain this, we perform simulation with one-color (760 nm) laser field, as shown in the Fig. \ref{fig:Figure4ATIandRS} (b). The oscillation of ATI spectrum dependent on CEP is clearly observed with one-color laser field at intensity of $1.9 \times 10^{14} W/cm^{2}$, as shown in Fig. \ref{fig:Figure4ATIandRS} (g), where there is no different between CEP=[0, $\pi$] and [$\pi$, $2\pi$]. The calculation results can be understood by the symmetry of electric field. For one-color laser field, there are more than one peaks to contribute the electron of trajectory 1, like A and B in positive direction or C and D in negative direction, as shown in Fig. \ref{fig:Figure4ATIandRS} (b). The interference between electrons of trajectory 1 in different optical cycles washes away the asymmetrical dependence of ATI spectrum on CEP. However, for the three-color laser field, there is only one dominate peak A in positive direction and two peaks B and C in negative direction, as shwon in Fig. \ref{fig:Figure4ATIandRS} (a). The amplitude of A is two time of amplitudes of D and E in the same direction; therefore, the three-color laser field is asymmetrical in terms of CEP and also for the electrons of trajectory 1. The comparison of ATI spectrum driven by one-color and three-color lasers give a clear picture on the contributions of electrons from two different trajectories. The interference of electrons from trajectory 2 causes the CEP dependent oscillation in ATI spectrum, furthermore the electrons from trajectory 1 break the symmetry for relative CEP= [0, $\pi$] and [$\pi$, $2\pi$] because of the asymmetrical electric field.

\begin{figure}[htbp]
\centering\includegraphics[width=12cm]{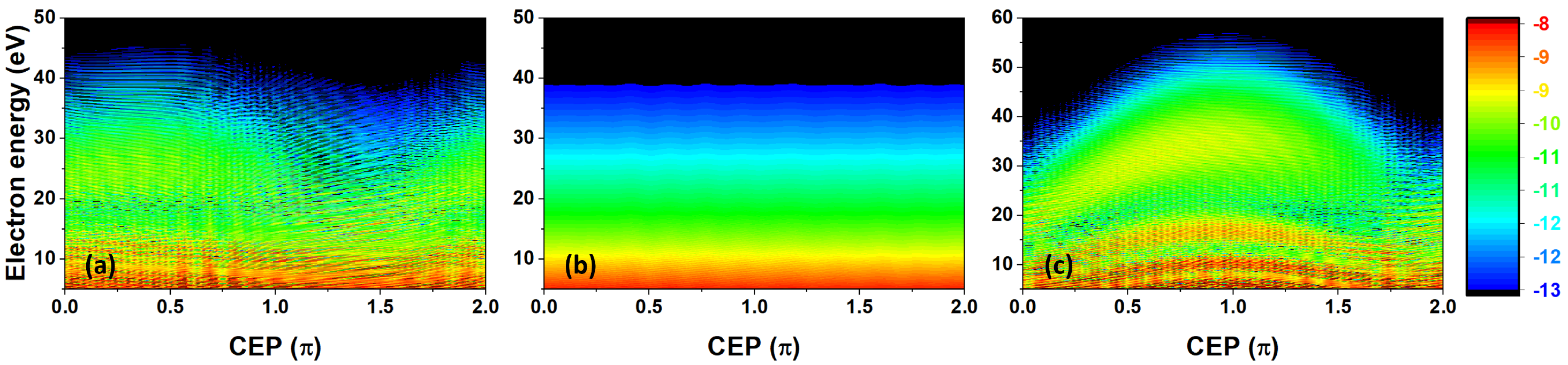}
\caption{Calculated ATI spectrum driven by three-color lasers with 3 optical cycles at intensity of $3.8 \times 10^{14} W/cm^{2}$ (a); by three-color laser with 2 optical cycles at intensity of $4 \times 10^{13} W/cm^{2}$ (b); by three-color laser with 2 optical cycles at intensity of $1.9 \times 10^{14} W/cm^{2}$ (c).}
\label{fig:Figure5MoreCaATI}
\end{figure}

In order to verify our understanding, the laser intensity is experimentally increased to $3.8 \times 10^{14} W/cm^{2}$ and it is found that the fast CEP-dependent oscillation disappears (see Supplemental material). The corresponding calculated ATI spectrum is shown in the Fig. \ref{fig:Figure5MoreCaATI} (a). It is found that there is no oscillation in ATI spectrum, where the DTI dominates and the FTI cannot significantly affect the ionization probability at this intensity. Furthermore, we theoretically investigate the effect of pulse duration to the ATI spectrum. When the pulse duration decreases from 7.8 fs (three optical cycles) to 5.2 fs (two optical cycles), the ATI spectrum changes quite a lot. At intensity of $4 \times 10^{13} W/cm^{2}$, the ATI spectrum is independent of CEP, as shown in Fig. \ref{fig:Figure5MoreCaATI} (b), which is similar to that of the Fig. \ref{fig:Figure4ATIandRS} (c). As the intensity increase to $1.9 \times 10^{14} W/cm^{2}$, there is no CEP-dependent modulation but a totally different structure in ATI spectrum, as shown in Fig. \ref{fig:Figure5MoreCaATI} (c). Our simulation results indicate that our experimental results are quite sensitive to the laser parameters, and fruitful physical phenomena may happen under different experimental conditions.

Last but not least, our experimental results show that our method provides a potential way to measure the absolute CEP with only one TOF. Calibrated by the numerical results, the absolute CEP of few-cycle laser pulse can be determined because of the asymmetry in the measured ATI spectrum. Compared with the traditional “stereo-ATI” experiment \cite{paulus2003measurement}, our results show a potential way to measure the absolute CEP of input laser with only one TOF, which is compact, easy to align and lower cost. In Ref. \cite{khurmi2017measuring}, C. Khurmi \emph{et al.} also propose a method to measure the absolute CEP with one TOF. Their method relays on that the Schrödinger equation of hydrogen atom in the strong laser field can be numerically solved with high precision. By comparing the experimental ATI spectrum from hydrogen atom with the simulation results, the absolute CEP is determined. However, with other noble gases, significant differences are observed between experimental data and simulation results. Compared with Ref. \cite{khurmi2017measuring}, our experimental data are obtained in Argon atoms, which may be provide a more versatile way to measure the absolute CEP. Currently our results only give a potential way to measure the absolute CEP, more experimental data and sophisticated simulation method are needed to improve our method into be more reliable and compact, where we can only apply the one-dimensional simulation because of the limitation of the computer at present.
\section{Summary}
In summary, ATI spectrum from Argon driven by few-cycle IR laser pulses is experimentally investigated when the Keldysh parameter is close to 1. For the first time, we observe a $\pi/5$ oscillation of whole ATI spectrum for relative CEP=[0, $\pi$]. Interestingly, this oscillation dramatically decreases for [$\pi$, $2\pi$] under our experimental conditions. The simulation results are well agreed with our experimental data quantitatively. We propose a semi-classical physical picture and contribute the experimental results to two kinds of electrons with different trajectories. The interference of electrons in FTI causes the fast oscillation while the electrons in DTI are responsible for the asymmetry in ATI spectrum. We propose a new experimental approach to modify the electric field to control the electron dynamics. Our results help to understand the physical mechanism of CEP modulated FTI process in strong laser field. Furthermore, the experimental results provide a potential way to determine the absolute CEP of few-cycle laser pulse with only one TOF spectrometer.
\section{Acknowledgement}
The authors thank Zenghu Chang for useful suggestions. B. B. Wang thanks Jing Chen for fruitful discussion. The work is finally supported by the National Natural Science Foundation of China (Grant No. 12034020, 12074418, 11774411, 11864037 and 91850209), and the Synergic Extreme Condition User Facility.

\bibliographystyle{alpha}
\bibliography{manuscript}

\newcommand{\etalchar}[1]{$^{#1}$}
\begin{thebibliography}{GRVDZN04}

\bibitem[AOIS21]{ayuso2021ultrafast}
David Ayuso, Andres~F Ordonez, Misha Ivanov, and Olga Smirnova.
\newblock Ultrafast optical rotation in chiral molecules with ultrashort and
  tightly focused beams.
\newblock {\em Optica}, 8(10):1243--1246, 2021.

\bibitem[BKI{\etalchar{+}}20]{boltaev2020application}
Ganjaboy~S Boltaev, Vyacheslav~V Kim, Mazhar Iqbal, Naveed~A Abbasi, Vadim~S
  Yalishev, Rashid~A Ganeev, and Ali~S Alnaser.
\newblock Application of 150 khz laser for high-order harmonic generation in
  different plasmas.
\newblock In {\em Photonics}, volume~7, page~66. Multidisciplinary Digital
  Publishing Institute, 2020.

\bibitem[CWC{\etalchar{+}}14]{chini2014coherent}
Michael Chini, Xiaowei Wang, Yan Cheng, He~Wang, Yi~Wu, Eric Cunningham,
  Peng-Cheng Li, John Heslar, Dmitry~A Telnov, Shih-I Chu, et~al.
\newblock Coherent phase-matched vuv generation by field-controlled bound
  states.
\newblock {\em Nature Photonics}, 8(6):437--441, 2014.

\bibitem[DMR{\etalchar{+}}09]{dunning2009engineering}
FB~Dunning, JJ~Mestayer, Carlos~O Reinhold, S~Yoshida, and J~Burgd{\"o}rfer.
\newblock Engineering atomic rydberg states with pulsed electric fields.
\newblock {\em Journal of Physics B: Atomic, Molecular and Optical Physics},
  42(2):022001, 2009.

\bibitem[EE14]{eilzer2014steering}
S~Eilzer and U~Eichmann.
\newblock Steering neutral atoms in strong laser fields.
\newblock {\em Journal of Physics B: Atomic, Molecular and Optical Physics},
  47(20):204014, 2014.

\bibitem[ENRS09]{eichmann2009acceleration}
Ulli Eichmann, T~Nubbemeyer, H~Rottke, and W~Sandner.
\newblock Acceleration of neutral atoms in strong short-pulse laser fields.
\newblock {\em Nature}, 461(7268):1261--1264, 2009.

\bibitem[ESE{\etalchar{+}}13]{eichmann2013observing}
U~Eichmann, A~Saenz, S~Eilzer, T~Nubbemeyer, and W~Sandner.
\newblock Observing rydberg atoms to survive intense laser fields.
\newblock {\em Physical review letters}, 110(20):203002, 2013.

\bibitem[GRVDZN04]{gurtler2004asymmetry}
A~G{\"u}rtler, F~Robicheaux, WJ~Van Der~Zande, and LD~Noordam.
\newblock Asymmetry in the strong-field ionization of rydberg atoms by
  few-cycle pulses.
\newblock {\em Physical review letters}, 92(3):033002, 2004.

\bibitem[HMS{\etalchar{+}}18]{hutten2018ultrafast}
Konrad H{\"u}tten, Michael Mittermair, Sebastian~O Stock, Randolf Beerwerth,
  Vahe Shirvanyan, Johann Riemensberger, Andreas Duensing, Rupert Heider,
  Martin~S Wagner, Alexander Guggenmos, et~al.
\newblock Ultrafast quantum control of ionization dynamics in krypton.
\newblock {\em Nature communications}, 9(1):1--5, 2018.

\bibitem[KWS{\etalchar{+}}17]{khurmi2017measuring}
Champak Khurmi, WC~Wallace, Satya Sainadh, IA~Ivanov, AS~Kheifets, XM~Tong,
  IV~Litvinyuk, RT~Sang, and D~Kielpinski.
\newblock Measuring laser carrier-envelope-phase effects in the noble gases
  with an atomic hydrogen calibration standard.
\newblock {\em Physical review A}, 96(1):013404, 2017.

\bibitem[KWZ{\etalchar{+}}21]{kubel2021high}
M~K{\"u}bel, P~Wustelt, Y~Zhang, S~Skruszewicz, D~Hoff, D~W{\"u}rzler, H~Kang,
  D~Zille, D~Adolph, GG~Paulus, et~al.
\newblock High-order phase-dependent asymmetry in the above-threshold
  ionization plateau.
\newblock {\em Physical Review Letters}, 126(11):113201, 2021.

\bibitem[LTM{\etalchar{+}}14]{li2014fine}
Qianguang Li, Xiao-Min Tong, Toru Morishita, Hui Wei, and Chii~Dong Lin.
\newblock Fine structures in the intensity dependence of excitation and
  ionization probabilities of hydrogen atoms in intense 800-nm laser pulses.
\newblock {\em Physical review A}, 89(2):023421, 2014.

\bibitem[LXH{\etalchar{+}}21]{liu2021electron}
Mingqing Liu, Songpo Xu, Shilin Hu, Wilhelm Becker, Wei Quan, Xiaojun Liu, and
  Jing Chen.
\newblock Electron dynamics in laser-driven atoms near the continuum threshold.
\newblock {\em Optica}, 8(6):765--770, 2021.

\bibitem[LZH{\etalchar{+}}20]{liu2020dynamics}
Jinlei Liu, Jing Zhao, Yindong Huang, Xiaowei Wang, and Zengxiu Zhao.
\newblock Dynamics of rydberg states and terahertz waves generated in strong
  few-cycle laser pulses.
\newblock {\em Physical Review A}, 102(2):023109, 2020.

\bibitem[LZZ{\etalchar{+}}16]{lv2016comparative}
Hang Lv, Wanlong Zuo, Lei Zhao, Haifeng Xu, Mingxing Jin, Dajun Ding, Shilin
  Hu, and Jing Chen.
\newblock Comparative study on atomic and molecular rydberg-state excitation in
  strong infrared laser fields.
\newblock {\em Physical Review A}, 93(3):033415, 2016.

\bibitem[MPBB06]{milovsevic2006above}
DB~Milo{\v{s}}evi{\'c}, GG~Paulus, D~Bauer, and W~Becker.
\newblock Above-threshold ionization by few-cycle pulses.
\newblock {\em Journal of Physics B: Atomic, Molecular and Optical Physics},
  39(14):R203, 2006.

\bibitem[NGS{\etalchar{+}}08]{nubbemeyer2008strong}
T~Nubbemeyer, K~Gorling, A~Saenz, U~Eichmann, and W~Sandner.
\newblock Strong-field tunneling without ionization.
\newblock {\em Physical review letters}, 101(23):233001, 2008.

\bibitem[NW06]{nakajima2006phase}
Takashi Nakajima and Shuntaro Watanabe.
\newblock Phase-dependent excitation and ionization in the multiphoton
  ionization regime.
\newblock {\em Optics letters}, 31(12):1920--1922, 2006.

\bibitem[PLW{\etalchar{+}}03]{paulus2003measurement}
Gerhard~Georg Paulus, Fabrizio Lindner, Herbert Walther, Andrius
  Baltu{\v{s}}ka, Eleftherios Goulielmakis, Matthias Lezius, and Ferenc Krausz.
\newblock Measurement of the phase of few-cycle laser pulses.
\newblock {\em Physical review letters}, 91(25):253004, 2003.

\bibitem[RHT{\etalchar{+}}06]{robinson2006generation}
JS~Robinson, CA~Haworth, H~Teng, RA~Smith, JP~Marangos, and JWG Tisch.
\newblock The generation of intense, transform-limited laser pulses with
  tunable duration from 6 to 30 fs in a differentially pumped hollow fibre.
\newblock {\em Applied Physics B}, 85(4):525--529, 2006.

\bibitem[RWC{\etalchar{+}}19]{ren2019line}
Xiaoming Ren, Yang Wang, Zenghu Chang, James Welch, Aaron Bernstein, Michael
  Downer, Jeffrey Brown, Mette Gaarde, Arnaud Couairon, Miroslav Kolesik,
  et~al.
\newblock In-line spectral interferometry in shortwave-infrared laser filaments
  in air.
\newblock {\em Physical review letters}, 123(22):223203, 2019.

\bibitem[VPT11]{volkova2011ionization}
EA~Volkova, AM~Popov, and OV~Tikhonova.
\newblock Ionization and stabilization of atoms in a high-intensity,
  low-frequency laser field.
\newblock {\em Journal of Experimental and Theoretical Physics},
  113(3):394--406, 2011.

\bibitem[WG21]{wang2021two}
Yimeng Wang and Chris~H Greene.
\newblock Two-photon above-threshold ionization of helium.
\newblock {\em Physical Review A}, 103(3):033103, 2021.

\bibitem[WLT{\etalchar{+}}16]{wang2016carrier}
Lifeng Wang, Xin Lu, Hao Teng, Tingting Xi, Shiyou Chen, Peng He, Xinkui He,
  and Zhiyi Wei.
\newblock Carrier-envelope phase-dependent electronic conductivity in an air
  filament driven by few-cycle laser pulses.
\newblock {\em Physical Review A}, 94(1):013827, 2016.

\bibitem[XLH{\etalchar{+}}20]{xu2020observation}
SongPo Xu, MingQing Liu, ShiLin Hu, Zheng Shu, Wei Quan, ZhiLei Xiao, Yu~Zhou,
  MingZheng Wei, Meng Zhao, RenPing Sun, et~al.
\newblock Observation of a transition in the dynamics of strong-field atomic
  excitation.
\newblock {\em Physical Review A}, 102(4):043104, 2020.

\bibitem[YHT{\etalchar{+}}14]{ye2014full}
Peng Ye, Xinkui He, Hao Teng, Minjie Zhan, Shiyang Zhong, Wei Zhang, Lifeng
  Wang, and Zhiyi Wei.
\newblock Full quantum trajectories resolved high-order harmonic generation.
\newblock {\em Physical Review Letters}, 113(7):073601, 2014.

\bibitem[YMH{\etalchar{+}}18]{yun2018coherent}
Hyeok Yun, Je~Hoi Mun, Sung~In Hwang, Seung~Beom Park, Igor~A Ivanov, Chang~Hee
  Nam, and Kyung~Taec Kim.
\newblock Coherent extreme-ultraviolet emission generated through frustrated
  tunnelling ionization.
\newblock {\em Nature Photonics}, 12(10):620--624, 2018.

\bibitem[ZHY{\etalchar{+}}13]{zhong2013effects}
Shiyang Zhong, Xinkui He, Peng Ye, Minjie Zhan, Hao Teng, and Zhiyi Wei.
\newblock Effects of driving laser jitter on the attosecond streaking
  measurement.
\newblock {\em Optics Express}, 21(15):17498--17504, 2013.

\bibitem[ZTY{\etalchar{+}}14]{zhang2014long}
Wei Zhang, Hao Teng, Chen-Xia Yun, Peng Ye, Min-Jie Zhan, Shi-Yang Zhong,
  Xin-Kui He, Li-Feng Wang, and Zhi-Yi Wei.
\newblock Long-term stabilization of carrier-envelope phase for few cycles ti:
  Sapphire laser amplifier.
\newblock {\em Chinese Physics Letters}, 31(8):084204, 2014.

\end{thebibliography}

\end{document}